\documentclass[aps,prl,reprint,superscriptaddress]{revtex4-1}

\usepackage{graphicx}
\usepackage{stmaryrd} % For \llbracket and \rrbracket
\usepackage{color}

\newcommand{\mom}[1]{\left\langle{#1}\right\rangle}
\newcommand{\moy}[1]{\left\langle{#1}\right\rangle}
\newcommand{\cum}[1]{\left\langle\!\left\langle{#1}\right\rangle\!\right\rangle}

\newcommand{\ilfrac}[2]{#1/#2}

\newcommand{\vac}{\ensuremath{V\!\!_{\text{ac}}}}
\newcommand{\vdc}{\ensuremath{V\!\!_{\text{dc}}}}
\newcommand{\fac}{\ensuremath{f\!_{\text{0}}}}

\newcommand{\fref}[1]{Fig.~\ref{#1}}
\newcommand{\eref}[1]{Eq.~(\ref{#1})}
\newcommand{\eqref}[1]{(\ref{#1})}
\newcommand{\eqs}{Eqs.~}
\renewcommand{\section}[1]{{\smallskip\em{#1}}.~}

\makeatletter
\newcommand\colorcapon{\renewcommand\fnum@figure{\figurename~\thefigure~(color online)}}
\newcommand\colorcapoff{\renewcommand\fnum@figure{\figurename~\thefigure}}
\makeatother

\begin{document}

\title{Photon pair shot noise in electron shot noise}
    \author{Jean Olivier Simoneau}
        \email{Jean.Olivier.Simoneau@USherbrooke.ca}
    \author{St\'{e}phane Virally}
    \author{Christian Lupien}
    \author{Bertrand Reulet}
    \affiliation{D\'epartement de Physique, Universit\'e de Sherbrooke, Sherbrooke, Qu\'ebec J1K 2R1, Canada}
\date{\today}

\begin{abstract}
%We report the measurement of the statistics of photons emitted by a tunnel junction that emits non-classical radiation. 
%This is obtained by measuring up to the fourth cumulant of the voltage fluctuations generated by the sample. 
%When the electron shot noise generates squeezed electromagnetic field, we provide a direct observation of the presence of photon pairs, as demonstrated by the Fano factor of the photon flux going above one.
%We report the measurement of the statistics of photons in the non-classical radiation emitted by a tunnel junction. 
%This is obtained by measuring up to the fourth cumulant of the voltage fluctuations generated by the sample. 
%When the electron shot noise generates a squeezed electromagnetic field, the measurement provides a direct observation of the presence of photon pairs, characterized by a Fano factor of the photon flux above unity.
%The former point of view, in its quantum version, is particularly adapted to describe the electron shot noise of a tunnel junction. 
There exists a fascinating dual representation of the electric ac current flowing through a normal conductor. 
On the one hand, it can be understood in terms of charge transport. 
On the other hand, it consists in an electomagnetic field guided by conducting structures embedded in an insulator. 
The former point of view, in its quantum version, is particularly adapted to describe the electron shot noise in a coherent conductor, like a tunnel junction at ultra-low temperature. 
However, when the junction is appropriately biased by a dc and an ac voltage, the noise it generates is best analyzed using the latter representation and the tools of quantum optics, as the radiation exhibits clear signs of non-classicality. 
Herein, we report the measurement of the statistics of photons emitted by such a tunnel junction. 
We observe a \emph{photon shot noise} characteristic of photon pair emission, as its Fano factor for small signal is above unity. 
The theory of \emph{electron shot noise}, dealing exclusively with the tunneling of charges through the junction, quantitatively fits the data from which photon shot noise is extracted. 
This experiment thus provides a clear link between the dual representations.
\end{abstract}


\maketitle

Electric ac current flowing through a mesoscopic sample at low temperature must be treated quantum mechanically and is described by an operator $\hat{I}$ instead of a classical time-dependent scalar $I(t)$. 
As a consequence, many properties of current fluctuations, which are characterized by current-current correlators such as $\langle\hat{I}(t)\hat{I}(t')\rangle$, cannot be explained by classical physics (for a review on current fluctuations, see \cite{blanter2000,nazarov2003}). 
This is particularly obvious in the short time correlations (i.e., high frequency noise), which directly exhibits the relevant time scales $h/k_BT$ and $h/eV$ associated with the temperature $T$ and voltage $V$ through Planck's constant, recently observed \cite{thibault2015}.
In the last years, the emphasis has shifted towards a different representation of electric transport, that of an electromagnetic field guided by the conducting structures to which the sample is connected. 
In such a representation, the current operator is replaced by the usual creation and annihilation operators of quantum optics.
Here again, quantum aspects become dominant at low temperature and high frequencies.

Recent experiments have underlined the non-classical nature of the electromagnetic field emitted by mesoscopic structures at low temperature, in particular using Josephson junctions between superconductors \cite{bozyigit2011, menzel2012, lang2013}. As with structures made of normal conductors, 
a tunnel junction under ac excitation has been shown to radiate a particular quantum state of the electromagnetic field, known as a squeezed state \cite{gasse2013}, where one quadrature of the field exhibits noise below the quantum limit, at the expense of added noise in the complementary quadrature. 
In the same system, strong correlations have been observed between the power fluctuations at two frequencies $f_1$ and $f_2$ verifying $f_1+f_2=f_0$, where $f_0$ is the excitation frequency \cite{forgues2014}. 
These measurements have been performed down to the single photon level, leading to the demonstration of photon-photon correlations that have long been known to exist in nonlinear quantum optics \cite{loudon2000quantum}. 
These observations have been explained by the presence of entanglement, in the light of the violation of Bell-like inequalities in the same system, shortly thereafter \cite{forgues2015}. 
Although quantum optics theory best explains these experiments in terms of a two-mode squeezed state \cite{loudon2000quantum}, it is remarkable that all the theory that has been used to fit the data quantitatively stems directly from the scattering matrix approach that only deals with electron quantum shot noise. 
%The link between the two representations has only very recently acquired some theoretical ground \cite{arne2016, mendes2015, beenaker2001, lebedev2010}.
The link between the two representations for normal conductors has been explored theoreticaly in various situations \cite{arne2016, mendes2015, beenaker2001, lebedev2010}.

Emission of photon pairs have clearly been demonstrated in the dual-frequency experiments, but it should also occurs in the single-mode squeezed state \cite{gasse2013}. 
In this case, since the frequencies of the emitted photon pairs are very close, the photon-photon correlation technique of \cite{forgues2015} cannot be applied. 
However, photocount statistics should reveal the signature of photon pairs. 
In particular, photon number resolving detectors would measure a Fano factor $\mathcal F = \ilfrac{\mom{\delta n^2}}{\mom{n}}$ of two at small average photon number $\mom{n}$, as is expected for a squeezed vacuum \cite{rosenberg2005, barnett1997methods}. 
Unfortunately, reliable photon detectors do not yet exist in the microwave domain, where the tunnel junction emits \cite{aguado2000, gustavsson2007, romero2009, chen2011, inomata2016}.
However, a direct link between the continuous statistics of voltage fluctuations and the discrete photocount distribution has recently been derived and tested with coherent radiation \cite{virally2016}. 
In this Letter, we use this technique to precisely demonstrate the existence of photon pairs in the radiation emitted by the tunnel junction.

This Letter is organized as follows. 
In a first section, we calculate the Fano factor of the photon flux generated by a tunnel junction in the presence of ac+dc excitation in conditions that correspond to our experimental data.
%the well-known theory of current-current correlators in mesosocpic conductors is combined with a newly published theory linking continuous current or voltage measurements with discrete photocount statistics. 
The experimental setup is presented in a second section followed by a description of the calibration procedure, in particular the effect of the saturation of the parametric amplifier which results in the distortion of the statistics of voltage fluctuations.
%along with a model that enables the recovery of actual continuous current statistics from a highly saturated amplified signal. 
The last section presents experimental results, including features of the discrete photocount statistics showing the presence of photon pairs emitted by the junction. 
It is followed by concluding remarks.

\section{Theory}
In order to demonstrate the existence of photon pairs in the radiation emitted by a conductor, we consider the average photon number $\moy{n}$ as well as its variance $\moy{\delta n^2}=\moy{n^2} - \moy{n}^2$. 
We define the Fano factor $\mathcal{F}=\mom{\delta n^2}/\mom{n}$, which has been demonstrated to move towards unity as $\mom{n}$ approaches zero, for all classical electromagnetic fields \cite{virally2016}. 
This is to be contrasted with non-unit values for purely quantum states. 
%In particular, it is expected that the Fano factor will move towards 2 if photons are emitted in pairs. 
In particular, the Fano factor is expected to reach 2 if photons are emitted in pairs. 
The photon number statistics is related to that of voltage fluctuations across the conductor. 
An intuitive link is given by the fact that the power per unit bandwidth of the radiation at frequency $f$ emitted by a conductor of resistance $R$ into an impedance matched detector can be expressed on the one hand as $R\,S_\mathrm{e}(f)/4$, and on the other hand as $\mom{n}hf$. 
Here, $S_\mathrm{e}(f)$ is the spectral density of current fluctuations at frequency $f$, where only the part that corresponds to photon emission has been kept. 
On the right hand side, $\moy{n}$ is the (unitless) average number of photons emitted per unit time and unit bandwidth, and $h$ is Planck's constant.

\begin{figure}
    \centering
    \includegraphics[width=.47\textwidth]{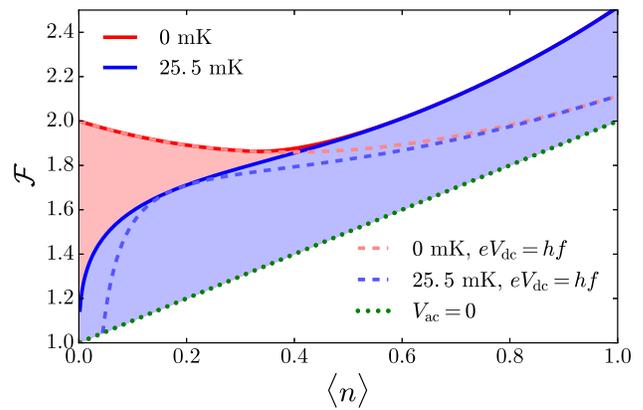}
    \colorcapon
    \caption{Theoretical Fano factor as a function of the average photon number for $T=0\,\textrm{mK}$ (red) and $T=25.5\,\textrm{mK}$ (blue). Dashed lines: fixed $\vdc$, varying $\vac$. Shaded areas correspond to $\vac$ and $\vdc$ ranging from $0$ to $3hf/e$. Solid lines: largest Fano factor. Dotted green line: $\vac=0$, varying $\vdc$. Calculations are done for $f=6\,\textrm{GHz}$ and $\fac=12\,\textrm{GHz}$.}
    \colorcapoff
    \label{FanoCaustics}
\end{figure}

To recover $\moy{n}$ and $\moy{\delta n^2}$, we use the results of a recent work that has rigorously made the general link between the discrete photocount statistics and the cumulants of a continuous voltage or current distribution \cite{virally2016}. 
They read
\begin{equation}
    \mom{n}=C_2-\frac{1}{2}
    %,
    \label{avgN}
\end{equation}
and
\begin{equation}
    \mom{\delta n^2}=\frac{2}{3}\,C_4+C_2^2-\frac{1}{4},
    \label{varN}
\end{equation}
with $C_n$ the $n\textsuperscript{th}$ cumulant of the continuous distribution \cite{mathstatone1951}. 
These cumulants have been calculated for a tunnel junction at low temperature using the theory of current statistics in coherent conductors \cite{blanter2000}.
When the sample is biased with a voltage $V(t)=\vdc+\vac\cos\left(2\pi \fac t\right)$, the so-called photo-assisted noise $C_2$ at frequency $f$ is
\begin{equation}
    C_2=\frac{R}{hf}\sum_{n=-\infty}^{+\infty}
    \frac{\alpha_n^2}{2} \left[S_0(f_{n+}) + S_0(f_{n-})\right],
    \label{eqC2}
\end{equation}
with $S_0(f') = \left(\ilfrac{hf'}{R}\right)\coth[hf'/(2k_BT_e)]$ the equilibrium noise spectral density at frequency $f'$ and temperature $T_e$. 
%In this equation, the noise spectral density has been rescaled to a number of photons per unit time and unit bandwidth. 
In addition, $\alpha_n \equiv J_n\left(\frac{e\vac}{h\fac}\right)$, where the $J_n$ are Bessel funtions of the first kind, $f_{n\pm} \equiv f+n\fac\pm e\vdc/h$, and $R$ is the resistance of the tunnel junction. 
With the same notations, the fourth cumulant $C_4$ is \cite{forgues2013, forgues2014}
\begin{equation}
    C_4=\frac32\left(\frac{R}{hf}
        \sum_{n=-\infty}^{+\infty}\frac{\alpha_n\alpha_{n+1}}{2}\left[S_0(f_{n+}) - S_0(f_{n-})\right]
        \right)^2.
    \label{eqC4}
\end{equation}
Combining these equations, we can calculate the Fano factor for any ac+dc voltage, frequency and temperature. 
In the absence of ac or dc excitation, $C_4$ vanishes. % and there are no photon pairs.  
As a result, $\mathcal{F}=\mom{n}+1$, which corresponds to thermal light, shown as a green dotted line in \fref{FanoCaustics}. 
Varying the dc and ac bias modifies both $C_2$ and $C_4$, thus $\mathcal{F}$. 
\fref{FanoCaustics} shows theoretical predictions for the Fano factor as a function of the average photon number, both at zero temperature (red) and at finite temperature $T_e=25.5\,\textrm{mK}$ (blue). 
Dashed lines correspond to a fixed dc voltage $\vdc=hf/e$ and varying the ac voltage. 
Spanning the full dc and ac biases yields the shaded areas. 
The full lines are the caustics corresponding to the largest Fano factor.
Calculations are done for $f=6\,\textrm{GHz}$, $\fac=12\,\textrm{GHz}$, with both $\vac$ and $\vdc$ amplitudes ranging from $0$ to $3\,hf/e$.
At zero temperature, the $\moy n=0$ limit is reached for $\vac=0$ and $\vdc$ bounded to $\left[ -hf/e ; hf/e \right]$. 
Under these conditions, we find $\mathcal F = \left[ \ilfrac{e\vdc}{(hf)} \right]^2+1$. 
%The  prediction for pure pairs is t
Thus, for $\vdc = \pm hf/e$, the Fano factor reaches its maximum of 2, a value corresponding to pure pair emission, as observed on the red curve of \fref{FanoCaustics}. 
At finite temperature, $\mom{n}>0$ even in the absence of bias. At very low bias, the Fano factor corresponds to that of thermal radiation,  %{\bf\color{red}(n can be 0, e.g. for $\vdc=0$ and $\vac=0$, or numerical error?)}
$\mathcal{F}\to1$, as witnessed on the blue line of \fref{FanoCaustics}. This is expected, as the radiation corresponds to that of a black body.
%More precisely, we find $\mathcal{F} = 1+1/\left[1+8hfk_BT_e/(e\vac)^2\right]$. 
%Once again, i
It is remarkable that the Fano factor, computed as a complicated combination of electron current correlators, has such a clear interpretation in terms of photons.

\begin{figure}[htb]
    \includegraphics[width=.97\columnwidth]{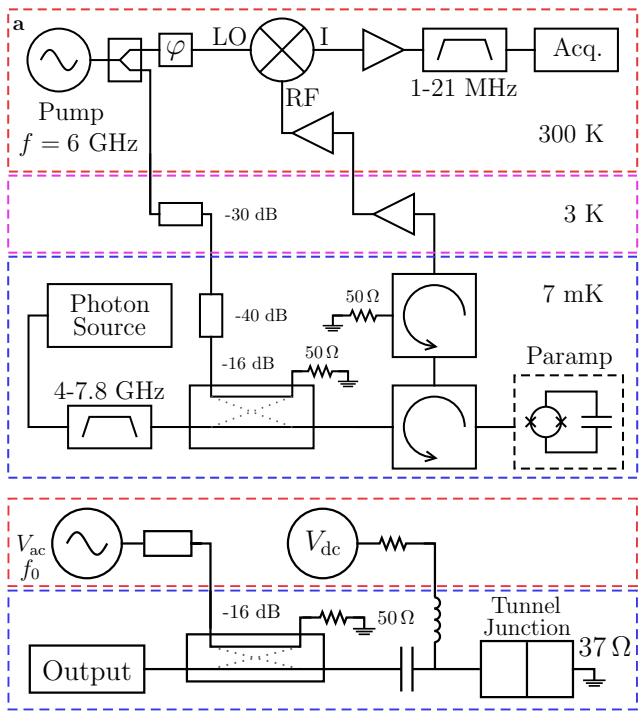}
    \colorcapon
    \caption{{\bf a)} Experimental setup used for detection. The signal emitted by a photon source is picked-up by an amplification chain with a paramp as its first link. It is then downconverted towards dc, using the same signal as the paramp pump, and digitized. {\bf b)} The photon source: a tunnel junction excited in ac+dc voltage.  (see text for more details).}
    \colorcapoff
    \label{Setup}
\end{figure}
\section{Experimental setup}
The experimental setup is presented in \fref{Setup}. 
We study the signal emitted by a photon souce (see \fref{Setup}b) made of an $\textrm{Al}/\textrm{Al}_2\textrm{O}_3/\textrm{Al}$ tunnel junction of resistance $R=37\;\Omega$ placed in a dilution refrigirator at 7~mK. 
A magnetic field ensures that the aluminum stays in the normal metal state. 
A bias-tee and a directional coupler allow for biasing the sample under both dc and ac voltage. 
The signal from the source is first amplified by a commercial nearly quantum-limited Josephson parametric amplifier (paramp) used in phase-sensitive mode and also cooled down to 7~mK. 
A 4-7.8~GHz band-pass filter limits the incident signal on the paramp. 
Circulators convey the pre-amplified signal to subsequent amplification stages at 3~K and 300~K, while isolating the paramp from additional amplification noise, thus keeping it close to the quantum-limited amplification regime. 
The amplification chain has a narrow passing band centered around 6~GHz, dictated by the frequency of the paramp pump. 
The amplified signal is down-converted by a mixer with a local oscillator (LO) at the same $f=6\,\textrm{GHz}$ frequency, and further filtered into a 1.2 to 21.4~MHz frequency band. 
The same source acts as paramp pump and mixer LO. The phase of the LO is controlled by a phase shifter, thus enabling the mixer to measure the amplified quadrature emitted by the paramp. 
Finally, the signal is sampled by a fast acquisition card with 14-bit resolution and a 400~MSa/s sampling rate. 
Measurement histograms are compiled on the fly, allowing the recovery of all cumulants of the voltage distribution.
Measurements are made for $\vdc$ values ranging from -56 to +56~$\mu$V and $\vac$ amplitudes spanning 40~dB.

\section{Calibration} 
In order to obtain a quantitative determination of the photon statistics emitted by the junction, it is necessary to calibrate the overall gain of the detection amplification $G$ as well as its noise power $T_A$.
%As usual, these parameters are estimated by measuring the variance of the current fluctuations and fitting the data with $G\left(\ilfrac{\left[S_0(f_{0+})+S_0(f_{0-})\right]}{2} +T_A\right)$. %, see \eref{S2}.
%A first estimate for these parameters is obtained by $G\left(\ilfrac{\frac{R}{4k_B}\left[S_0(f_{0+})+S_0(f_{0-})\right]}{2} +T_A\right)$, as usual.
A first estimate for these parameters is obtained by measuring the shot noise of the junction as a function of $\vdc$, as usual \cite{spietz2003}.
From these measurements we estimate %$T_A=2.489$~photons and an electron temperature of $T_e=27\,\textrm{mK}$. 
$T_A\simeq 717\,\textrm{mK}$, which corresponds to $\simeq2.5$ added photons, and an electron temperature of $T_e\simeq27\,\textrm{mK}$.
In order to calibrate the ac voltage experienced by the sample, we measured the photo-assisted noise and fitted it with \eref{eqC2}. 
The very low noise power of the setup is obtained by using an almost quantum limited parametric amplifier. 
There is however a drawback to the use of this device: it is nonlinear, even at the level of the tiny signals we are measuring. As a consequence, the distribution function of the current fluctuations is altered by the paramp. 
While the variance of the fluctuations is only slightly modified (by about 1\%, see below), higher order cumulants are strongly polluted. 
To correct for the distortion induced by the paramp, we relate the digitized signal $\Sigma(t)$ to the voltage $s(t)$ generated by the sample with the following model:
%model it as follows:
\begin{equation}
    \Sigma=\sqrt{G}\left[(s+v)+\alpha(s+v)^3+w/\sqrt{g}\right]
    .
    \label{eq_SatModel}
\end{equation}
Here, %$s$ is the signal generated by the sample, $\Sigma$ the signal we detect, 
$\alpha$ represents the effect of the non-linearity ($\alpha<0$), $v$ is the noise added by the paramp at its input, and thus that will also experience distortion, and $w$ is the noise of the setup that is added after the paramp. 
$g$ is the power gain of the paramp and $G$ the total power gain of the amplification chain. 
From the $n\textsuperscript{th}$ measured cumulant $K_n=\cum{\Sigma^n}$, we wish to recover the $n\textsuperscript{th}$ cumulant of the original signal $\cum{s^n}$ which, after proper rescaling, provides $C_n$.
The usual calibration procedure from the variance $K_2$ provides an estimation of $G$ and $\cum{w^2/g+v^2}$. 
In order to find the other parameters, we have measured the fourth cumulant $K_4$ in a situation where we know that the fluctuations emitted are almost Gaussian, so that $\cum{s^4}\ll \cum{s^2}^2$. 
This happens in the absence of photo-excitation. 
In this case, the intrinsic non-Gaussian fluctuations we expect are indeed extremely small, in particular thanks to the narrow bandwidth of our detection.
\begin{figure}[htb]
    \centering
    \includegraphics[width=.47\textwidth]{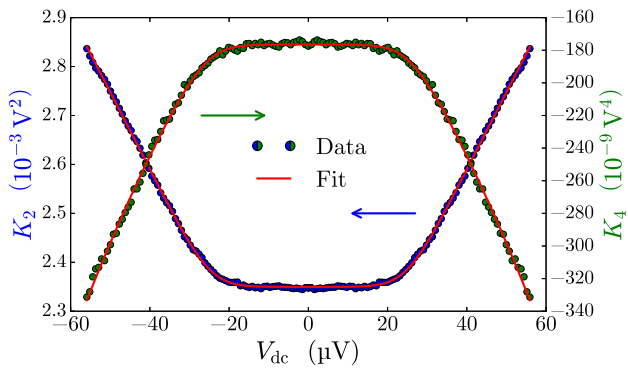}
    \colorcapon
    \caption{Least squares fit results used for the calibration of the saturation model. $K_2$ and $ K_4$ are fitted simultaneously to the \eref{eq_SatModel} model using a unique parameter set.}
    \colorcapoff
    \label{Fit}
\end{figure}
%As a consequence, the measured $K_4$ detected for $\vac$ is only due to the nonlinearity of the detection. 
As a consequence, the observed $K_4(\vac=0)$ is non-zero only because of the nonlinearity. 

In \fref{Fit} we show the measurements of both $K_2$ and $K_4$ as a function of $\vdc$ for $\vac=0$. 
The data are very well fitted by the model, from which we deduce a paramp noise power of $620\,\textrm{mK}$ (corresponding to $2.16$ added photons), while the rest of the amplifiers contribute $87\,\textrm{mK}$ (corresponding to $0.30$ added photon) to the total noise, and an electronic temperature of $T_e=25.5\,\textrm{mK}$. 
%Those results are very close to the values estimated by disregarding the saturation.
The obtained total amplification noise temperature as well as the electronic temperature are very close to the values estimated by disregarding the saturation.
The parameter $\alpha=-6.11\times10^{-7}$ corresponds to a cubic non-linearity, usually referred to as IP3 in the language of amplifiers. 
Here, $\textrm{IP3}\simeq-165\,\textrm{dBm}$. 
Note that we have not included any $s^2$ term in our model (i.e., no IP2). 
This comes from the fact that, contrary to many semiconducting devices, the non-linearity of the Josephson junction $I=I_c\sin\varphi$ is an odd function of the phase, i.e. positive or negative voltages are equivalent. 
We checked that a quadratic non-linearity does not reproduce at all the observed $K_4$. 
In the following, we measure $K_2$ and $K_4$, from which we deduce $C_2$ and $C_4$ using the saturation model of \eref{eq_SatModel}.

\section{Results}
In order to probe the behavior of the photocount distribution in various situations, measurements are performed for two distinct ac frequencies, $\fac=11.4$~GHz and $\fac=12$~GHz. 
%Indeed, we expect the source to only emit pairs of photons for which the sum of frequencies is equal to the excitation frequency \cite{forgues2014}. %, as dictated by conservation of energy \cite{forgues2014}. 
Indeed, the tunnel junction emit pairs of photons at any frequencies $f_1$ and $f_2$ such that $f_1+f_2 = \fac$  \cite{forgues2014}. %, as dictated by conservation of energy \cite{forgues2014}. 
Since we only detect photons in a narrow bandwidth $[f-\delta f,f+\delta f]$ around $f=6\,\textrm{GHz}$, both $f_1$ and $f_2$ must fall within this interval (i.e. $|f_{1,2}-f|<\delta f=21$ MHz) for photon pairs to be observed. 
Thus, if we excite at $11.4\,\textrm{GHz}$ no photon pair can be detected. 
In contrast, if we excite at 12~GHz, we do expect to measure photon pairs.
%Hence, in the case when we excite the source at 11.4~GHz and measure photons in a narrow band around 6~GHz, we do not expect to measure any signature of photon pairs. 
%If one photon is emitted around 6~GHz, we expect its counterpart in the pair to be mitted around 5.4~GHz, well outside of our measurement window. 

%Results of both experiments (for $\fac=11.4$~GHz and $\fac=12$~GHz) are presented in \fref{C2C4} and \fref{F}. %
\begin{figure}[htb]
    \centering
    \includegraphics[width=.47\textwidth]{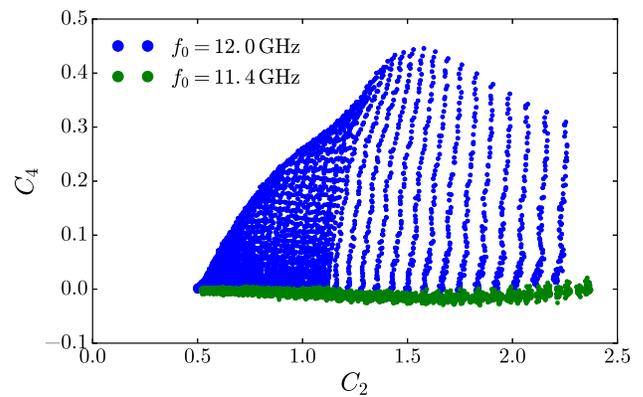}
    \colorcapon
    \caption{Second and fourth-order cumulants of current fluctuations generated by the tunnel junction, measured at frequency $f=6$ GHz, for all $\vdc$, $\vac$ and $\fac$ combinations.}
    \colorcapoff
    \label{C2C4}
\end{figure}
\begin{figure}[htb]
    \centering
    \includegraphics[width=.47\textwidth]{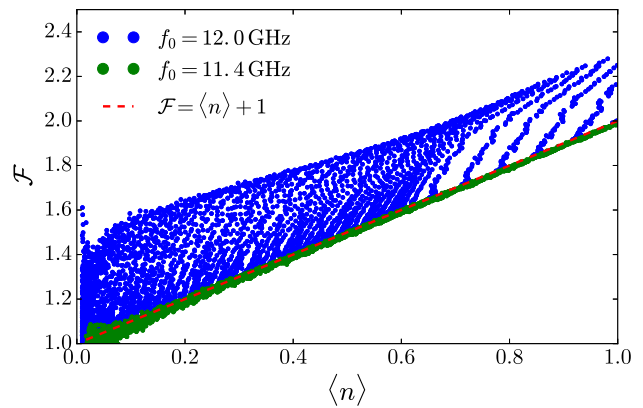}
    \colorcapon
    \caption{Fano factor as a function of the average photon number.}
    \colorcapoff
    \label{F}
\end{figure}
For both excitation frequencies, we have measured $C_2$ and $C_4$ as a function of $\vdc$ and $\vac$. 
Instead of presenting the results for both $C_2(\vdc,\vac)$ and $C_4(\vdc,\vac)$, a more insightful representation of the data is provided by plotting $C_4$ as a function of $C_2$ for the full $(\vdc,\vac)$ parameter space, as we do in \fref{C2C4}. 
In this figure, green dots represent the results for $\fac=11.4$~GHz. 
%{\bf\color{red} Discuss $C_2$ and $C_4$ in their entirety before discussing the fano factor/photon statistics.}
We clearly observe that $C_4$ vanishes in this case. The $C_4$ detected around frequency $f$ is indeed proportional to the current--current correlator $\mom{I(f)^2}$, which is nonzero only if $2f=pf_0$ with $p$ integer \cite{gasse2013}. 
In other words, this mismatched photo-excitation does not modify the quasi-Gaussian nature of the narrow band electron shot noise at frequency $f$.
In the case of an excitation at twice the measured frequency ($\fac=12\,\textrm{GHz}$, blue dots), $C_4$ no longer vanishes for nonzero $\vdc$ and $\vac$.
This is consistent with the results of \cite{gasse2013}, where squeezing is observed under the same photoexcitation conditions, as well as with \cite{forgues2013} for the classical regime.
In this case, the current statistics is non-Gaussian. 
%Orange dots represent a particular data subset for $\vdc\approx hf$. %, which would be the optimal situation for $T_e=0\,\textrm{mK}$.

From $C_2$ and $C_4$, we use \eqs\eqref{avgN} and \eqref{varN} to present $\mathcal F$ vs. $\mom n$ in \fref{F}.
In the $\fac = 11.4\,\textrm{GHz}$ case, observed photons are unpaired and the statistics is known to be thermal \cite{barnett1997methods}.
In view of \eqs\eqref{avgN} and \eqref{varN}, this explains the vanishing $C_4$ as, for thermal light, $\mom{\delta n^2}=\mom{n}\!\left(\mom{n}+1\right)$.
This situation corresponds to the green dots that are constrained onto $\mathcal F = \mom n +1$ (dashed red line).
In the $\fac = 12\,\textrm{GHz}$ case, we do observe the signature of photon pairs as an increased Fano factor (blue dots). 
As shown in \cite{virally2016}, in the $\mom n \to0$ limit, the Fano factor tends towards unity for all classical signals.
However, it can deviate from that value for quantum signals such as that generated by a squeezed state. 
For a pure squeezed vacuum, a signal purely made of pairs, it would actually reach two.
We do not reach this ideal value in this experiment because the finite temperature mixes the squeezed vacuum with a weak thermal state. 
Those experimental results compare favorably with the theory shown in \fref{FanoCaustics} for the actual experimental electronic temperature of $25.5\,\textrm{mK}$.

\section{Conclusion}
By measuring the statistics of the current fluctuations generated by a quantum conductor, we have been able to deduce the photon statistics of the eletromagnetic field it radiates. 
In particular, we have provided a direct observation of the shot noise of photon pairs generated by electron shot noise, as evidenced by the observation of the Fano factor of the photon flux. 
It is noticeable that, as far as we know, such an observation has never been reported even in conventional quantum optics because of the lack of photon number resolving detectors. 
Our experiment opens new perspectives in the study of non-classical microwave radiations, in particular for signals with no phase reference, where complete tomography is irrelevant. One example is the radiation of Josephson junctions in high impedance environments, which have recently attracted much attention \cite{armour2015,grimm2015}. 

\vspace{\baselineskip}
We thank G.~Lalibert\'e for technical help and S. Boutin for fruitful discussions.
This work was supported by the Canada Excellence Research Chairs, Government of Canada, Natural Sciences and Engineering Research Council of Canada, Qu\'ebec MEIE, Qu\'ebec FRQNT via INTRIQ, Universit\'e de Sherbrooke via EPIQ, and Canada Foundation for Innovation.
%This work was supported by the Canada Excellence Research Chairs program, Canada NSERC, Qu\'{e}bec MEIE, Qu\'{e}bec FRQNT via INTRIQ, Universit\'{e} de Sherbrooke via EPIQ, and Canada Foundation for Innovation.

\bibliography{article}

\end{document}